\documentclass[twocolumn, amsmath,amssymb, superscriptaddress]{revtex4}

\usepackage{dcolumn}% Align table columns on decimal point
\usepackage{bm}% bold math
\usepackage{amsmath}
\usepackage{amsfonts}
\usepackage{graphics}
\usepackage[pdftex]{graphicx}
\usepackage{times}
\usepackage{setspace}
\usepackage{verbatim}
\usepackage[raggedright]{titlesec}
\usepackage{color}   % {\color{red} TEXT }

\begin{document}

\title{The $Z$-index: A geometric representation of productivity and impact which accounts for information in the entire rank-citation profile}

\author{Alexander M. Petersen}
\address{IMT Lucca Institute for Advanced Studies,  55100 Lucca, Italy} 
\author {Sauro Succi}
\address{Istituto Applicazioni Calcolo C.N.R.,  Rome, Italy} 
\address{Freiburg Institute for Advanced Studies, Albertstrasse, 19, D-79104, Freiburg, Germany} 

\begin{abstract}
We present a simple  generalization of Hirsch's $h$-index, $Z \equiv \sqrt{h^{2}+C}/\sqrt 5$, where $C$ is the total number of citations. 
$Z$ is aimed at correcting  the potentially excessive penalty made by $h$ on a scientist's 
highly cited papers,  because for the majority of scientists analyzed, we find the excess citation fraction $(C-h^{2})/C$  to be distributed closely  around the value 0.75, meaning that 75 percent of the author's impact is neglected. Additionally, $Z$ is less sensitive to local changes in a scientist's citation profile, namely perturbations  which increase $h$ while only marginally affecting $C$.
Using real career data for 476 physicists careers and 
488 biologist careers, we analyze both the distribution of $Z$ and the rank stability of $Z$ with respect to the Hirsch index $h$ and the Egghe index $g$. We analyze careers distributed across a wide range of total impact, including top-cited physicists and biologists for benchmark comparison. 
In practice, the  $Z$-index requires  the same information 
needed to calculate $h$ and could be effortlessly  incorporated within career profile 
databases, such as {\it Google Scholar} and {\it ResearcherID}. Because $Z$  incorporates information from the entire publication profile while being more robust than $h$ and $g$ to local perturbations, we argue that $Z$ is better suited for ranking comparisons in academic  decision-making scenarios comprising a large number of scientists. 
\end{abstract}

\maketitle 

\section{Introduction}
%\footnotetext[1]{ Send correspondence to petersen.xander@gmail.com or succi@iac.cnr.it}
The most commonly used quantitative measure of a scientist's publication portfolio is
Hirsch's $h$-index, which was designed to measure both productivity and impact simultaneously \cite{H}. 
However, there have been 
many criticisms claiming that the  $h$-index leads to inconsistencies in scientific ranking, represents productivity 
or impact but not both, is non-decreasing and hence cannot be used as a short-term evaluation metric, and oversimplifies the publication portfolio   \cite{hindexResearchers,ProConH,ComparisonIndex,hindexFields,hindexQ,Halternative,37hvariantes,hInconsistency}.
As  such, despite its increasing use in diverse decision-making processes,  it may not be optimally suited 
for career evaluation scenarios. By way of example, it has recently been implemented by the  
National Agency for the Evaluation of Universities and Research Institutes (ANVUR) of Italy as selection and
pay-scale criterion in the most recent  large-scale national ``habilitation'' competition \cite{ANVUR}.
Furthermore, without proper normalization of  $c$, the number of citations to a given paper, in order to account for variations across 
time, group size, academic discipline, and even academic sub-discipline, the practice of comparing citation 
counts without normalizing is highly questionable \cite{UnivCite,Rad2,Scientists,hindexResearchers,hindexFields,FracCounts,CiteBiases}. 

\begin{figure*}[t]
\centering{\includegraphics[width=0.99\textwidth]{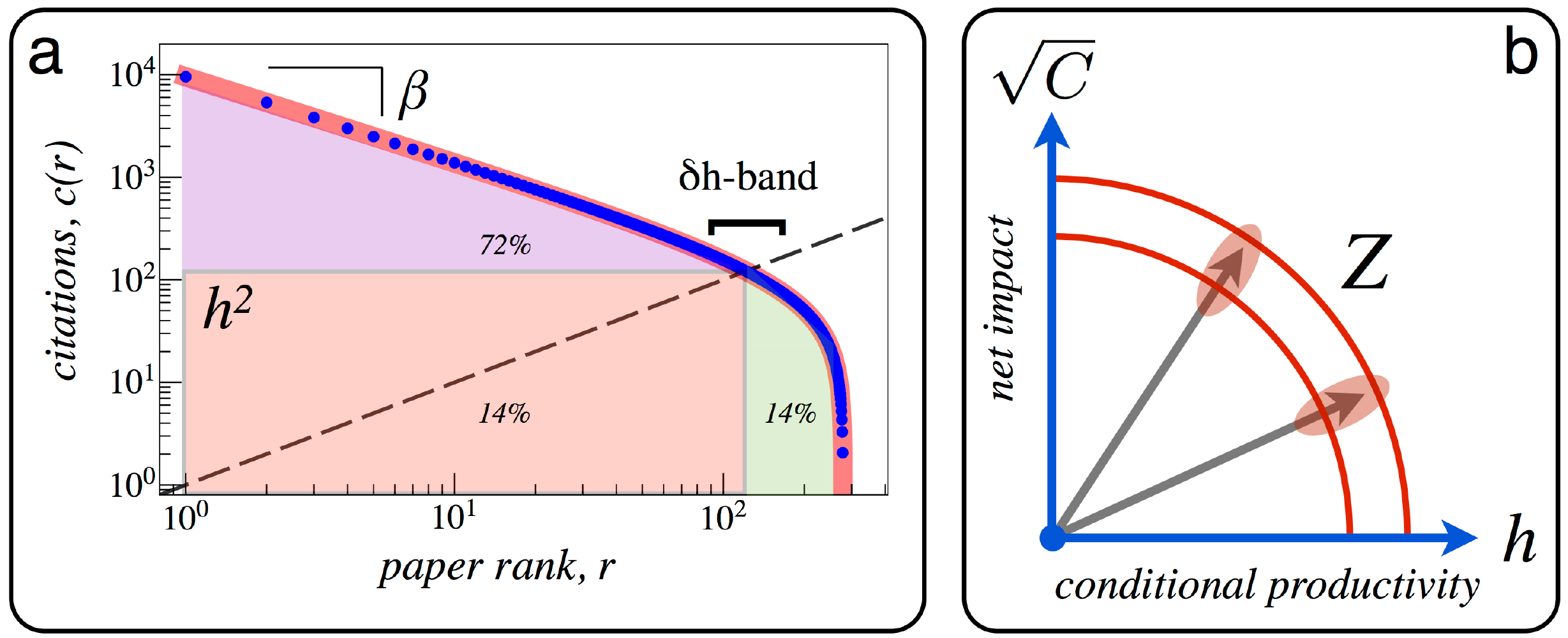}}
\caption{\label{p1} Understanding the relation between $C$ and $h$. (a) Schematic illustration of the  rank citation profile $c_{i}(r)$, illustrated as a discrete generalized beta distribution (DGBD) defined in Eq. (\ref{Cr}) using $N=278$, $\beta_{i} \equiv 0.83$, $\gamma_{i} \equiv 0.67$, $A_{i} = 220$, $C_{i}= 79,342$, and $h_{i} = 104$ (reproduced from the schematic example in \citealt{RankCitSciRep}). All single-valued quantitative indicators can be calculated from $c_{i}(r)$. The citation count $h^{2}$ (shaded pink) accounts for only a subset of the total citations $C$. In fact, for careers similar to the rank-citation profile shown, where the top papers are roughly distributed according to a power law $c_{i}(r)\sim r^{-\beta_{i}}$, then $h_{i}^{2}$ represents only a tiny fraction of the total citations $C$. 
Indeed, in this example which is characteristic of prestigious  scientists, the excess citations of the peak papers contribute to the  majority of the total $C$. 
The ``$\delta h$ band'' is the set of papers with rank $r \in [h \pm \delta h]$ that, once cited,  optimally 
increase the $h$-index, while only marginally affecting $C$. For this reason $h$ is unstable with respect to local perturbations in $c(r)$.
Because $Z$ is a geometric combination of $h$ and $\sqrt C$, the quantity $Z$  is less sensitive to local changes in a scientist's citation profile, and hence less susceptible to direct manipulation.
 (b) Visualizing the $z$-plane. The radial isolines for constant $Z= \sqrt{h^{2} + C}/\sqrt{5}$ enclose careers  
 with similar $Z$ values. Given the diversity of careers, and the variability in $c(r)$, these isolines capture 
the two-dimensional features of both conditional (relatively high-impact) productivity as captured by $h$, and 
net impact of all publications as measured  by $C$.}
\end{figure*}

The Hirsch index integer  $h$ counts the number of publications in a scientists's portfolio which satisfy the criterion of having $h$ or more citations each. 
A scientist's rank-citation profile,
 $c_{i}(r)$, is calculated by ranking the $N_{i}$ total publications of a given scientist $i$  in decreasing order of 
 citations, so that $c_{i}(1) \geq c_{i}(2) \geq \dots \geq c_{i}(N_{i})$.
The  significance threshold $h$ is chosen somewhat arbitrarily by using the ``fixed point'' relation, 
corresponding to  $c_{i}(h_{i})=h_{i}$, which is most easy to communicate by graphically visualizing 
the  rank-citation distribution. 
Fig. \ref{p1}(a) illustrates how $h$ corresponds to the intersection of the line
$c=r$ with  $c_{i}(r)$.
The mathematical definition of $h$ is that the paper of rank $h$ has no less than $h$ citations:
\begin{equation}
 c_{i}(h_{i}) \geq h_{i} \ , 
\end{equation}
which insignificantly alters the graphical definition$^{1}$. 
\footnotetext[1]{ For the purpose of brevity and compactness, we shall tend to suppress the author index $i$ from 
career measures throughout the rest of the paper, unless explicit reference to the author dependence is crucial.}

However, the arbitrary choice of the quality threshold $h$, whereby papers with only $h$ or more citations are counted, implicitly
makes this more a measure of {\it productivity} conditioned to a quality threshold. 
Hence, a second measure of impact is required, since an obvious feature of a 2-dimensional representation 
is that there be at least two independent degrees of freedom.
Here, we propose the two-dimensional representation of productivity and impact in the $z$-plane defined by the coordinates $(h, \sqrt{C})$, which incorporates the total citations $C = \sum_{r} c(r)$ as a complementary impact measure to  $h$, which is principally a productivity measure. 

Moreover, the simple geometric combination 
\begin{equation}
Z \equiv \frac{\sqrt{h^{2}+C}}{\sqrt 5}
\end{equation}
computed by using the information typically provided in online career profile databases and better incorporates
the non-trivial information contained in the entirety of a scientist's rank-citation profile $c_{i}(r)$. 
In what follows, using analytic and empirical demonstration based on real data for 964 scientists, we will show 
several basic properties of $Z$ which preserve the conveniences of $h$ while gaining 
robustness to perturbations in $c(r)$ and also providing a better representation of the net impact of the {\it entire} rank-citation profile.

\section{Why yet another index?}

The utility of the $h$-index is that it provides a remarkably calculable, easy-to-memorize, and comparable 
quantitative summary of the information contained in the full rank-citation distribution $c(r)$. 
It captures the difficulty in science of consistently producing highly-cited papers, and discourages 
voluminous publication strategies which may lack overall quality. 

The appreciation of quality over quantity is an issue that must be emphasized 
as science makes a revolutionary shift away from old
system of publication based on peer-review in printed  journals with fixed capacity towards a 
rapid and unlimited online capacity system of the near future. In this future scenario, measures similar to the $h$-index will be important for extracting the signal from the noise, and set quality thresholds 
by which to measure productivity.

However, the $h$-index comes with a number of well-known weaknesses  \cite{hindexResearchers,ProConH,ComparisonIndex,hindexFields,hindexQ,Halternative,37hvariantes,hInconsistency}.
Below we provide a partial list of those that are most relevant to the motivation for $Z$:
\begin{enumerate}
\item The quality threshold $c^*$ used to highlight the representative papers for which $c(r) \geq c^*$, is 
arbitrarily chosen to be $c^*\equiv h$, corresponding to the fixed point solution of $c(h) \geq h$. Other arbitrary significance thresholds are used in alternative productivity-impact measures, such as the ``i10-index'', the number of publications with at least 10 citations, which is listed on {\it Google Scholar} profiles. It is poorly understood how the choice of the significance threshold $c^*$ may alter the overall distribution of the impact indicator across scientists, and whether or not there is a ``best'' choice for $c^*$.
\item The $h$-index severely discounts the impact of the highly cited papers, for which $c(r)\gg h$, in a scientist's publication portfolio. This point is particularly important, since in science, like in sports, many notable distinctions are awarded to recognize top performance rather than overall continued impact. 
\item Motivated by the competitive reward system, scientists may begin to adapt strategies that ``game the system'' of prestige. According to the definition of the $h$-index as a single point on $c(r)$, it is indeed possible for a scientist who has accurate knowledge of his/her $c(r)$, to selectively self-cite his/her papers in the ``$\delta h$-band''  $r \in \{h-\delta h, h+ \delta h\}$, in order to 
optimally increase his/her $h$ index in the near future (see Fig. \ref{p1}(a)). 
\end{enumerate} 

These criticisms are not new, and many alternative metrics have been proposed to mend these weaknesses \cite{Halternative,37hvariantes}.
Among others, the  Egghe  $g$-index, defined by $g^2 \le \sum_{r=1}^g c(r)$, is designed to provide more
weight to the highly cited papers within $c(r)$ \cite{G}. Alternatively, it can
more readily be appreciated as a fixed point measure of the average number of citations calculated for the first $g$ papers, 
$g \le \sum_{r=1}^g c(r)/g = \langle c \rangle_{g} $. However, as we will show in the later section, $g$ and $h$ are quantitatively related and highly correlated, and hence do not measure remarkably different information contained in $c(r)$.
 The neglected citation count  $C-h^2$ mostly belong to the highly-cited papers, and are the motivation for a complementary excess $e$-index \cite{E}. 
Figure \ref{p1}(a) shows the rank-citation profile of a typical highly-cited scientist with a significant number of highly cited papers. Indeed,  the neglected citations $C-h^2$ account in this case for 86\% of the total citations!
Nevertheless, the $h$-index is popular because it sends a quick and efficient reputation signal that most practicing scientists can readily appreciate relative to their  peers. 
 
\section{Data analyzed}
 Here we analyze real career data for 476 physicist careers and 488 biologist careers. 
Each dataset contains careers distributed across a wide range of total impact, including the 
top-100  physicists and top-100 biologists according to total citation counts in high-impact journals. These two datasets
 will be used as an elite benchmark.
We use (i)  disambiguated  ``distinct author'' data from {\it Thomson Reuters Web of Knowledge} (TRWOK), {\tt www.isiknowledge.com/}, using 
their matching algorithms to identify publication profiles of distinct authors, and (ii) scientist profiles from ResearcherID.com, the open portal for TRWOK which allows individuals to aggregate publications into an online  repository which conveniently calculates both $h$ and $C$. \\

For the selection of two comparison sets for high-impact physicists, we aggregate all authors who published in {\it Physical Review Letters} {\it (PRL)} over the 50-year period 1958-2008 into a common dataset. From this dataset, we rank the scientists using the citations shares metric defined in 
\cite{Scientists}, and choose the top 100 scientists resulting in  dataset [A] (average $h$-index $\langle h \rangle \pm$ Std.Dev. $ = 61 \pm 21$).
As a comparative set of highly cited physicists, we also choose from our ranked {\it PRL} list, approximately 
randomly, 100 additional highly prolific physicists comprising dataset [B] (average $h$-index $\langle h \rangle = 44 \pm 15$). 
We compare the tenured scientists in datasets A and B with 100 relatively young assistant professors from physics in dataset [C]  ($\langle h \rangle = 14 \pm 7$). 
To select dataset [C] scientists, we chose two assistant professors from the top 50 U.S. physics and astronomy departments, ranked  according to the magazine {\it U.S. News}. Further analysis of the publication trajectories and collaboration patterns of physicists in dataset [A,B,C] is provided in \cite{RankCitSciRep,GrowthCareers,Reputation}.
Dataset [D] is comprised of  174 ``graphene'' scientists with profiles on {\it ResearcherID.com}, ($\langle h \rangle = 15 \pm 11$). 
Dataset [E] ($\langle h \rangle = 92 \pm 35$) is comprised of the top 100 scientists who published in the journal {\it Cell}, using the same ranking method as with dataset [A].  Datasets [F], [G] , and [H] correspond to scientists with profiles on {\it ResearcherID.com} with the keywords ``molecular biology'' ($\langle h \rangle = 20 \pm 17$), ``neuroscience'' ($\langle h \rangle = 17 \pm 13$), and ``genomics'' ($\langle h \rangle = 18 \pm 14$),  comprising 60, 76, and 252 profiles, respectively. Only ResearcherID profiles  with more than 7 publications were analyzed.

\section{Empirical Results}
\subsection{Accounting for highly-cited papers by rescaling the $h$-index}

\begin{figure*}[t]
\centering{\includegraphics[width=0.99\textwidth]{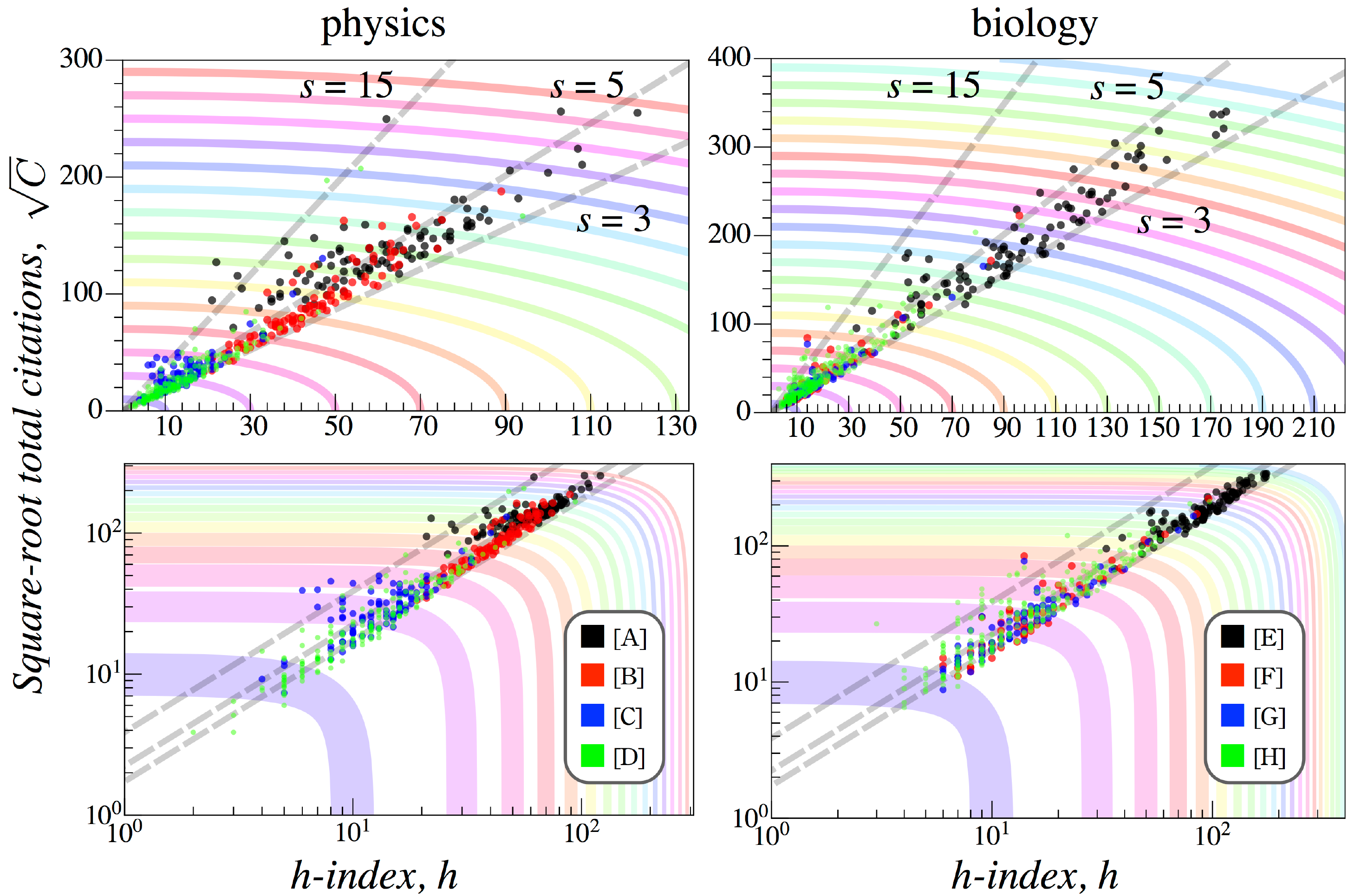}}
\caption{\label{p3} Scatterplot of empirical  $h,\sqrt C$ values in the $z$-plane.
The colored curves are constant $Z$ isolines of exponentially decreasing thickness with increasing $Z$. The straight dashed lines correspond
to the rescale factors with slope  $R(s)$  using   $s=3,5,15$.  The middle 50\% of the careers  are enclosed by the lines with $s=3$ and 5 (since the quartiles $Q_{0.75} = 5.2$ and  $Q_{0.25} = 3.3$). Interestingly, the low-$h$ outliers consisting of many assistant professor profiles (blue data points) tend to be well described by clusters along $Z$ isolines, suggesting that assistant professors are typically hired using criteria that select for relatively large $C$. }
\end{figure*}

A main weakness of the
$h$-index, its severe neglect of a scientist's highly cited papers, can be remedied by a two-dimensional representation 
of the net impact $\sqrt{C}$ and conditional productivity $h$.
We propose a  
representation of authors  by the data pair $(h,\sqrt{C})$ in a two-dimensional  $z-$plane shown in Fig. \ref{p1}.
We define the index $Z \equiv \sqrt{C+h^{2}}/ \sqrt{5}$, which is simply a vector norm of the coordinate $(h,\sqrt{C})$ using the ``natural'' units corresponding to $h/\sqrt{5}$ and $C/5$. The factor of $5$ is  
chosen according to statistically robust patterns between $h$ and $C$ which we discuss next.

Since $h^{2}$ is a subset of the citations calculated by $C$, then these two values are  
highly correlated,
\begin{equation}
C_{i} = s_{i} h_{i}^2 \ ,
\label{EqS}
\end{equation}   
where  this slope parameter $s_{i} \ge 1$, was noted originally by Hirsch (noted as $a$ \cite{H}). 
Recently, it was shown empirically by S. Redner \cite{Hredner} that the distribution of $s$ across 
authors is highly peaked around $s \approx 4$.
As a result, one may conclude that there is little point of combining
two highly correlated indicators. 
Indeed, the quantity $\sqrt{C}$ is the $h$-index an individual would have if all 
of his/her $N_{i}$ papers had the same number of citations, hence a completely flat $c(r)$. 
In the following, we shall argue oppositely; namely that the little
spread in the slope can nevertheless lead to sensible 
readjustments between ``peaky" and ``flat" authors, who deviate
significantly away from the characteristic value $s=4$.    
For this reason, we define $Z$ using the ``natural units'' factor $1/\sqrt{5}$, so that the rescaling of $h$ is mild 
in the region around $s=4$, but becomes fairly substantial 
in the tails of the $s$ distribution.

Figure \ref{p3} shows the scatter plots of $(h_{i},\sqrt{C_{i}})$ pairs for each scientist $i$ on linear and log-log axes. 
Colored bands represent isolines of constant $Z$.
The data are confined to a relatively small radial bands of the phase space,
mainly because $h$ is highly correlated with $C$ \cite{Hredner,RankCitSciRep}.
The fact that data are almost entirely collected between the rescaling lines with slope $Z(s)/h =$ 5 and 15, indicates the utility of the polar representation of  $(h_{i},\sqrt{C_{i}})$. The angle defining each datapoint is proportional to $s$, and is not as informative as the overall magnitude 
$Z$, which, we argue, captures a great deal of the summary information for each career.

\begin{figure*}[t]
\centering{\includegraphics[width=0.99\textwidth]{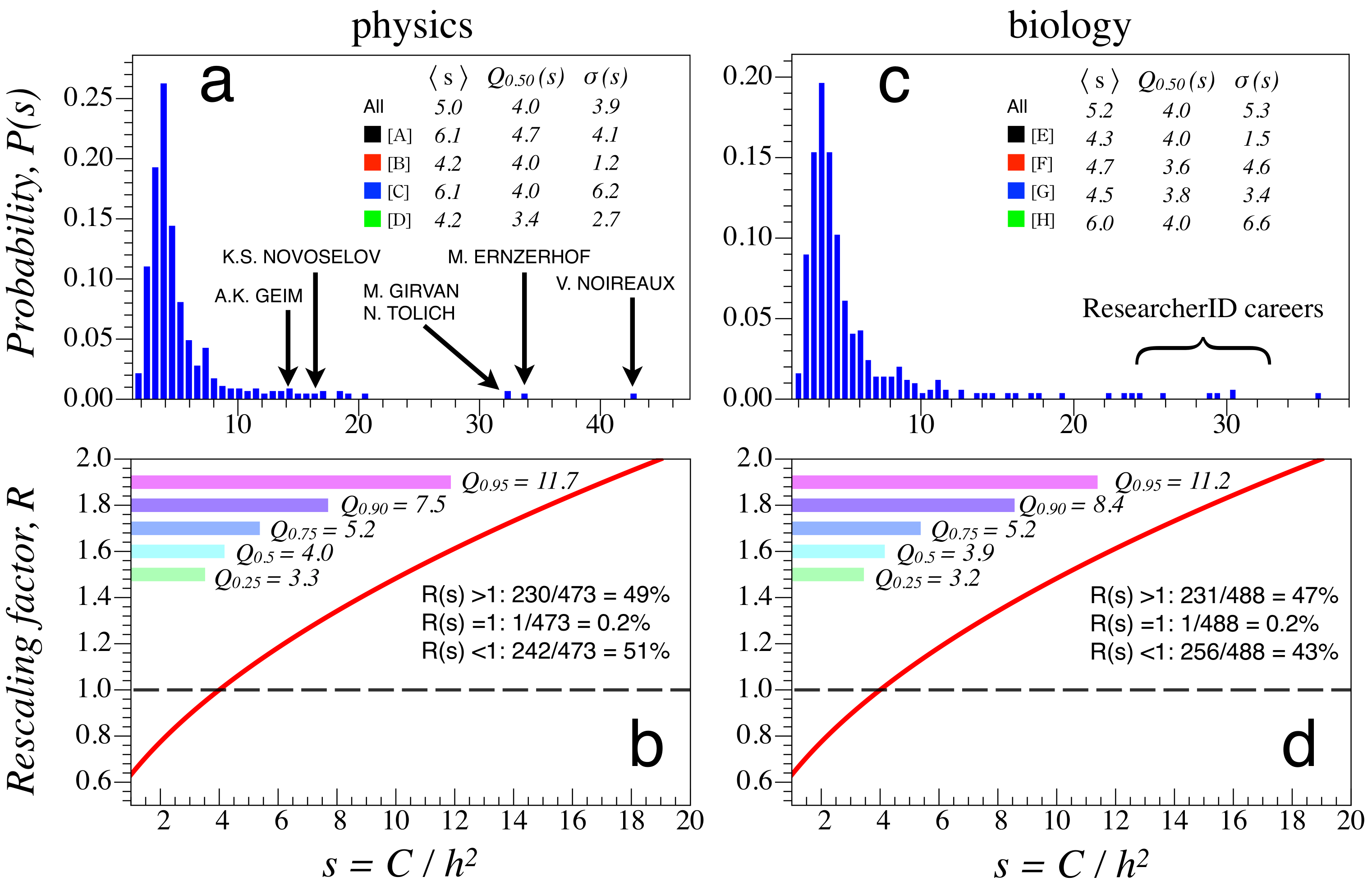}}
\caption{\label{p2}  Rescaling the $h$-index. (top) Empirical distribution of the scale-factor $s$ calculated for all careers within
each disciplinary field aggregated into a single dataset.  In panels $a,c$, each  distribution $P(s)$ is peaked around $Median(s) = 4.0$. Mean, standard deviation, and median values are also listed for each individual subset [A]-[D].  Standout careers, such as the recent Nobel Prize in Physics winners KS Novoselov and AK Geim, have large $s$ values due to their seminal publications. Other extreme $s_{i}$ values can arise if the number of publications $N_{i}$ is not considerably large and so $C_{i}$ is dominated by just a few high-impact papers. (bottom) The rescale factor $R(s) = \sqrt{(1+s)/5}$ is a slow function of $s$ around unity for $s=4$. Bar lines indicate the quantile $Q_{x}$ calculated from the empirical cumulative distribution of $s$ values. For instance in panel $b$,  90\% of the careers analyzed have $s < 7.5$, and the middle 50\% have  values in the range $3.3<s<5.2$.  Shown are the number and \% of careers in each dataset with rescale value $R(s)>1$, $R(s)=1$, and $R(s)<1$.}
\end{figure*}

Figure \ref{p2} shows  the probability distribution  $P(s)$ which  is peaked around the median $Q_{0.5}(s) = 4.0$.
This means that, for the majority of scientists analyzed, the excess citation fraction $(C-h^{2})/C = 1-1/s$  is distributed closely  around the value 0.75, meaning that $75\%$ of a career's citation impact is neglected by $h$.
 The standard deviation $\sigma(s)$  is largely dataset dependent, ranging from 1.2 to 6.6 due to the potential for extremely large $s$ values arising from careers with a large citation difference between the top-cited paper(s) and the rest  of the papers. 
Nevertheless,  most careers are  contained within a relatively narrow radial band in the $z$-plane. 

Because of the regularities in the statistical distribution of $s$, we define the norm using the ``natural units'' for $Z$, leading to the formulation
\begin{equation}
\label{bigH}
Z = h \sqrt{\frac{1+s}{5}} \ .
\end{equation}    
The ``natural units'' normalization factor $1/\sqrt 5$ means that $Z$ retains  the advantages of the scalar index $h$ since 
careers with $s=4$ correspond to the traditional $h$-index,   $Z(s=4)=h$.

So what is gained by using $Z$ instead of $h$?
Mainly,  $Z$ does not discount the value of significantly cited papers, those papers 
from which a scientist derives much of his/her scientific reputation \cite{Reputation}.
A second practical advantage of $Z$ 
is its robustness against  perturbations in the $z$-plane, $(h,\sqrt{C}) \to (h+1,\sqrt{C+1})$.
Such  perturbations could arise from just the stochastic inflow of citations, or possibly from covert ``cosmetic surgery''
self-citation strategies, aimed at increasing the $h$-index.

 We calculate the perturbation to $Z$, 
\begin{equation}
\delta Z = \frac{1}{\sqrt 5} \Big( \sqrt {(h+1)^2 + C+1} - \sqrt {h^2 + C} \Big)   \ ,
\end{equation}
resulting from a citation landing perfectly on a paper with $h$ citations located  in the center of the  ``$h$-band'' (see Fig. \ref{p1}).
Hence, $\delta Z \sim \frac{h+1}{5 Z} \sim 1/5 \ll 1$ for profiles with $Z \approx h$.
More generally, from the definition, the change of $Z$ in the
generic transition $(h,C) \to (h+\delta h, C+\delta C)$ is given by
\begin{equation}
\delta Z = \frac{1}{5 Z } (h \delta h + \frac{1}{2}\delta C)
\end{equation}
Assuming $Z  \sim h$, this expression shows that, in order to
increase $Z $ by one unit, the change $\delta h=1$ must
be accompanied by a change in the total citations of the order of
$\delta C \sim 8 Z $. In other words, $C$ serves as an
inertial reservoir of citations, preventing rapid changes due 
to local adjustments in $c_{i}(r)$. 
In summary, a single citation raising $h$ by one unit
(not a marginal increment, since for a productive scientist $h$ grows 
by approximately one unit a year) would have no effect on $Z$, rendering covert self-citation strategies less rewarding.

In the lower panels of Fig. \ref{p2} we also plot the rescaling factor 
\begin{equation}
R(s) = \frac{Z(s)}{h} =\sqrt{\frac{1+s}{5}}
\end{equation} 
between $Z$ and $h$, and show the fraction $x$ of scientists in each dataset having $R$ less than various quantile values $Q_{x}$.
$R(s)$  is a slow monotonically increasing function of $s$, demonstrated 
by the perturbation  $s=4+\Delta$, which varies the rescaling factor as  $R \sim 1 + \Delta/10$
for $\Delta \ll 1$.
Hence, authors with $\Delta \ll 10$ receive 
a mild (linear) correction to their $h$-index, while outliers at both extremes 
may be significantly affected by the rescaling. 
For example, Fig.~\ref{p2}(b) indicates that the middle 50\%  of physics careers analyzed
have $s$ values between 3.3  and 5.2, corresponding to  $R(3.3) = 0.92$ and $R(5.2) = 1.11$, respectively.

\subsection{Accounting for the entire rank-citation profile $c(r)$}
It was recently shown, for a large range of careers, that the {\it entire} citation profile $c_{i}(r)$ can be 
quantified  with a relatively simple parametric class of rank distribution model, the discrete generalized beta distribution (DGBD) 
\begin{equation}
c_{i}(r) \equiv A_{i} r^{-\beta_{i}} (N_{i}+1-r)^{\gamma_{i}} \ .
\label{Cr}
\end{equation}
The DGBD is well-suited for systems with finite number of constituents, as shown recently for rank-ordering of systems in the arts and sciences \cite{DGBfunc,RankOrder,RankCitSciRep,Reputation}. The $\beta$ parameter controls the logarithmic ``Zipf-law'' slope for small $r$ (high rank) constituents, whereas the $\gamma$ parameter controls the ``exponential-like'' cutoff for large $r$ (low rank) constituents.

A schematic example of a characteristic DGBD $c_{i}(r)$ of an elite scientist is plotted in Fig. \ref{p1}(a).
With only three degrees of freedom, this distribution is able to capture $c_{i}(r)$ across the entire range of $r$, as demonstrated for a broad range of careers in physics, including even assistant professor careers  with $N_{i} \sim 20$.
It remains an open problem whether scientists can be robustly classified in terms of the parameters $\beta_{i}$, $A_{i}$, and $C_{i}$.

It is possible to approximate the coefficient $A_{i}$ in Eq.~(\ref{Cr}) using the fixed-point definition $c(h)\equiv h$, which implies that
$A/h^{\beta} \approx h$. 
Hence, from Eq.~(\ref{Cr}) it follows that the expected total number of citations  can be approximated knowing $\beta$ and $h$  for a given profile by 
\begin{equation}
C_{\beta, h} \equiv \sum_{r=1}^{N} A r^{-\beta} \approx  h^{1+\beta} \sum_{r=1}^{N} r^{-\beta} = h^{1+\beta}H_{N,
\beta} \ .
\end{equation}
  Since the  {\it generalized harmonic number}   $H_{N, \beta}$ is on the  order  $O(1)$ for $\beta \approx 1$, then we arrive at the simple scaling relation $C_{\beta, h} \sim h^{1+\beta}$ \cite{RankCitSciRep} which agrees with empirical findings that $C\sim h^{2}$ \cite{Hredner} in the typical case  where $\beta \approx 1$. A recent study of an extremely large dataset of more than 30,000 $c_{i}(r)$ profiles \cite{RadH} indicates that the global distribution of $\beta$ values is indeed more concentrated with $\beta > 1$ values, corresponding to $c_{i}(r)$ that have a sharp decay from the highest cited paper to the bulk of the papers, representative of the majority of non-elite scientists. This is in stark contrast to the   $\beta \leq 1$ values  found for  top scientists reflecting the slow decay in $c_{i}(r)$ due to a large subset  of highly-cited papers.
  
  \begin{figure*}[t]
\centering{\includegraphics[width=0.79\textwidth]{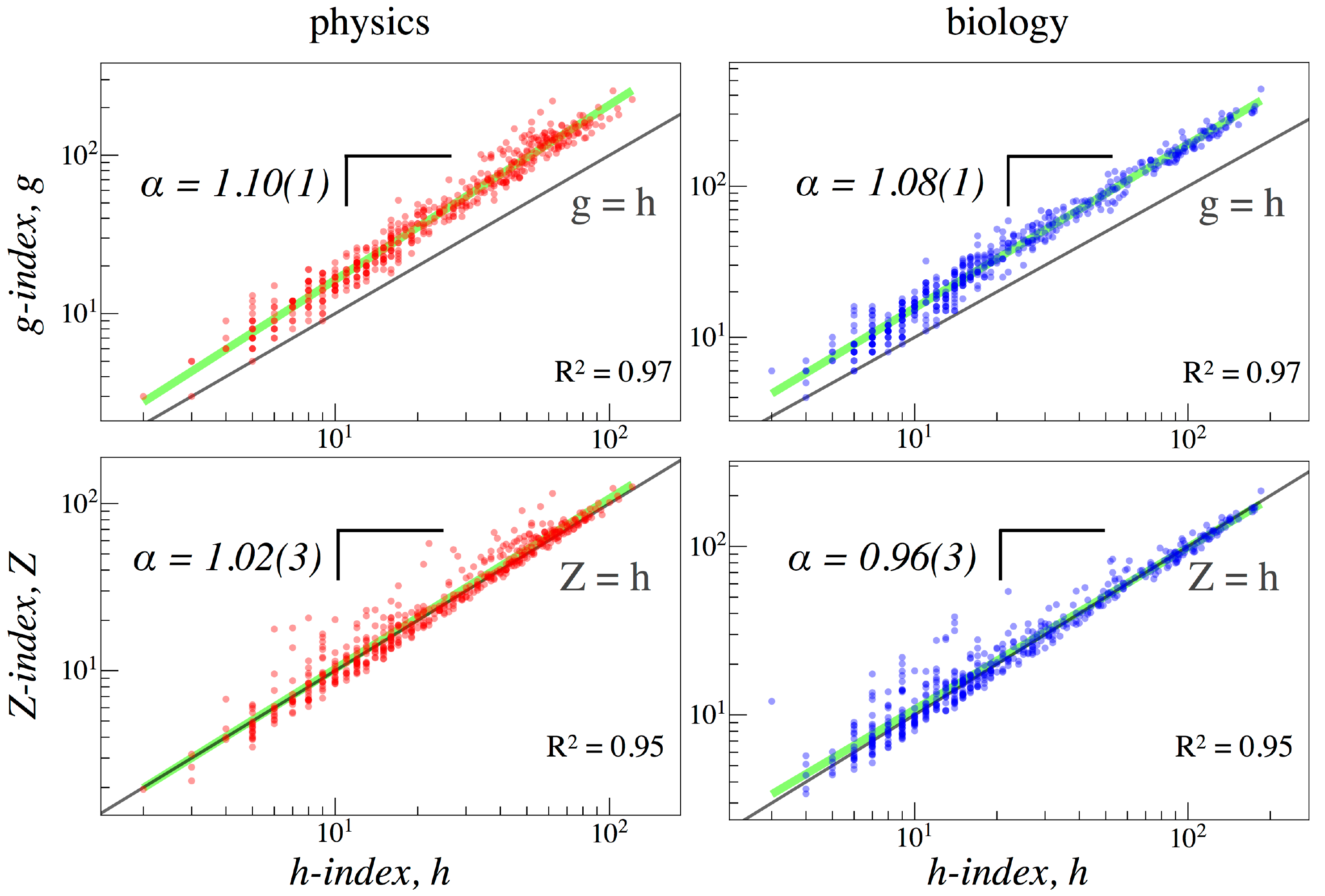}}
\caption{\label{p5} $h$ and $g$ are nearly linearly related. (Upper panels) Regression of $h$ and $g$ values indicate that these to measures are somewhat redundant. This feature follows from the relation $g_{i} \sim h_{i}$ for DGBD $c_{i}(r)$ with $\beta_{i} \approx 1$.
The high value of the ordinary least squares regression $R^{2}$ indicates that the information contained in $h$ is roughly equivalent to the information contained in $g$. (Lower panels) By construction, $h$ and $Z$ are also approximately linearly related. However, the presence of larger deviations from the $Z=h$ line, as indicated by slightly lower $R^{2}$ values, indicates that for certain profiles, the inclusion of $C$ is crucial to account for the entire rank-citation profile.}
\end{figure*}
  
Furthermore, a similar calculation shows that  $g^{2}\sim h^{1+\beta}$, and  hence $g\sim h$  for $\beta \approx 1$. 
In these heuristic calculations we neglect the $\gamma_{i}$ scaling regime since the low-rank papers typically contribute a relatively small amount to the overall $C_{i}$ tally as compared to the high-rank papers.

Together, these simple scaling relations show that $h$ and $g$ are closely related, conveying roughly the same overall information. 
We test this prediction empirically  in the upper panels of Fig. \ref{p5} which shows  for each disciplinary set that $g\sim h^{\alpha}$ with $\alpha \approx 1.10 \pm0.01$ for the physicists and $\alpha \approx 1.08 \pm 0.01$ for the biologists. Deviations from unity arise since the $\beta_{i}$ scaling exponent  is not universal, but varies around $\beta \sim 1$.  In the lower panels of Fig. \ref{p5} we show the analogous scatter plot and power-law model regression between $Z$ and $h$, which show slightly smaller $R^{2}$, indicative of certain profiles that are extremely penalized according to $h$, but which are compensated by $Z$.
Since $\beta_{i}$ values  are significantly more difficult to calculate than $h_{i}$, requiring regression or maximum likelihood calculations,   
we propose that the  simple two-dimensional data pair $(h,\sqrt{C})$ conveniently and sufficiently capture  scientific production-impact profiles across a broad range of age and prestige.

By way of example, consider the following physics careers with various $\beta_{i}$ distinguishing
how ``steep'' each $c_{i}(r)$ is in the highly-cited regime.
The average  $\beta$ value calculated across datasets [A,B,C] is $ \langle \beta \rangle \approx 0.77$. 
Hence, scientists with $\beta=0.78$, similar to A. H. Castro Neto with
   $(h=30,C=3509)$ corresponding to $s=3.9$ and $R=0.99$, would not be impacted, as $Z = h$.
Scientists with relatively large $\beta=1.04 $, such as R. B. Laughlin with
$(h=32,C=7751)$ corresponding to $s=7.6$ and $R=1.3$, would be significantly upgraded to $Z = 42$.
A relatively flat profile with $\beta \sim 0.5$ and $(h=30,C=2000)$ corresponding to $s=2.2$ and $R=0.8$, would be 
downgraded to $Z  =24$.
For a pathologically extreme case, consider E. Lieberman (not included in datasets [A-D]), with $(N=h=6, C=2530)$  corresponding to $s=70$, $R=3.8$, and $Z=23$. Alternatively, 
consider the  ``perfectly flat'' author with $(N=h=30,C=h^{2})$ corresponding to $s=1$, $R=\sqrt{2/5}$, and $Z = 19$. 
We believe that these readjustments represent a fair reward to the highly-cited peak papers 
which are discounted when considering $h$ alone.

\begin{figure*}[t]
\centering{\includegraphics[width=0.99\textwidth]{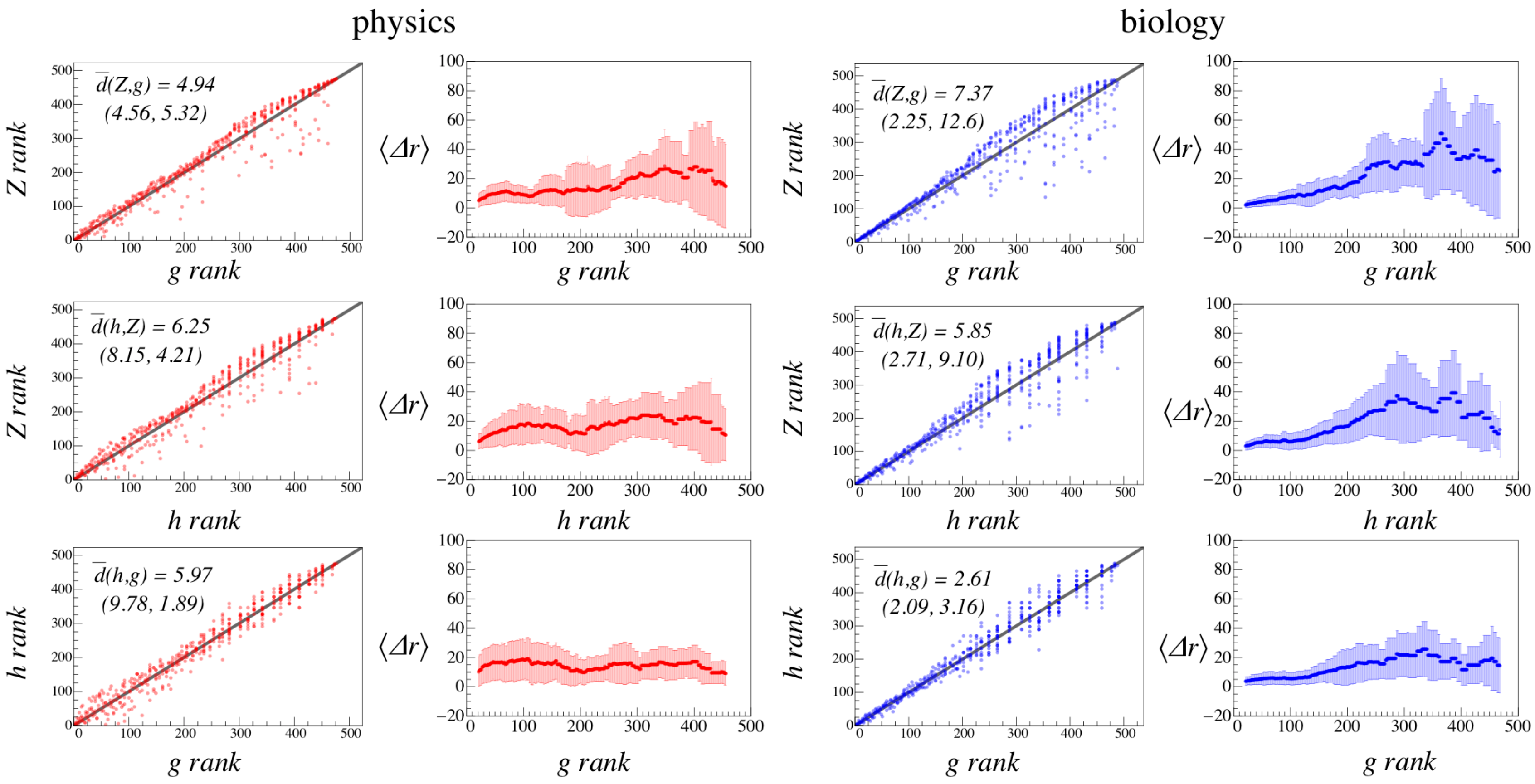}}
\caption{\label{p4}  Quantifying the rank stability of different indices.
Panels in columns 1 and 3 show the scatter plots of  rank pairs $(r_{i}(A),r_{i}(B))$ calculated using the three combinations of rankings according to  $g_{i}$, $h_{i}$, and $Z_{i}$ calculated for each scientist $i$. Panels in columns 2 and 4 show the corresponding rank-change distance $\Delta r $ as a function of the rank value $r_{i}(A)$. Shown are the running averages $\langle \Delta r \rangle$ calculated over centered bins of width $2\delta r +1 \equiv 21$ with 1-standard deviation error bars which measure the variations in rank stability. For the index $B=Z$ in the top two rows, there is an increasing rank instability for increasing $g$ and $h$ values, indicating that the $Z$ incorporates information that is neglected, resulting in significant shifts in the  overall ranking.   We also list the normalized Kullback-Leibler relative entropy $\overline{d}(r_{i}(A),r_{i}(B))$  defined in Eq.~(\ref{NKL}) for each data sample, and in parentheses list  the pair $(\overline{d}_{-}, \overline{d}_{+})$  calculated over the lower half and the upper half of each data sample.}
\end{figure*}

\subsection{Testing the rank-stability of $h$, $g$, and $Z$}
The rank stability of a system, either with respect to time evolution or variation in the ranking variable, is largely influenced by both the amplitude and the diffusive properties  of the noise in the system \cite{rankstability}. 
In the case of ranking of scientists according to quantitative measured derived from $c_{i}(r)$, it is important to quantify 
the rank stability of two comparative indices, say index $A$ and index $B$. We expect that  there will be more rank stability when the
two measures being compared  representing the same information derived from $c_{i}(r)$. However, if there is new information entering into
index $A$ that is not incorporated into index $B$, then we expect there to be larger fluctuations in the rankings of the scientists. 

We test this underlying assumption using three indexes, $h$, $g$, and $Z$, and compare the rank-stability of each pair. 
For each disciplinary dataset, we calculate the rank of each scientist, $r_{i}(A)$ according to each index, $h$, $g$, and $Z$. We then compare the rank of each scientist $r_{i}(B)$ according to a different index $B$, resulting in 3 scatter  plots for each disciplinary dataset, as shown in Fig. \ref{p4}.  

We quantify the rank-stability of each index pair $(A,B)$ using two methods. First,  for the rank pair $(r_{i}(A),r_{i}(B))$ for each career $i$, we calculate the 
distance 
\begin{equation}
\Delta r = \vert r_{i}(A)-r_{i}(B)\vert / \sqrt{2}
\end{equation} 
from the line $y=x$. The line $y=x$ is the stability benchmark corresponding to the ideal case of perfect rank stability, $r_{i}(A) = r_{i}(B)$. In order to investigate the relation between $\Delta r$ and $r$, Fig. \ref{p4} shows the running averages $\langle \Delta r \rangle$, with 1-standard deviation error bars. Specifically, the $\langle \Delta r \rangle$ is averaged over rank intervals $I_{r} \equiv [r-\delta r, r+\delta r]$ using $\delta r = 10$ and shown are the running average for $r=[(1+\delta r)...(S-\delta r)]$. The overall trends are not dependent on the choice of $\delta r$.

For the pair $A=g$ and $B=h$ the values of $\langle \Delta r \rangle$ are relatively small over the entire range of $r_{g}$, reflecting the property that $g\sim h$, and hence incorporate roughly the same information from $c_{i}(r)$. However, comparing the ranking for $A=h$ and $A=g$ to $B=Z$ we find that the amplitude of $\langle \Delta r \rangle$ significantly increases for large $r_{h}$ and $r_{g}$, reflecting an increasing rank-instability for the scientists with relatively small publication portfolios, and hence, more ``noise'' in their productivity-impact scores when measured by $h$ and $g$. This underscores the importance of using quantitative measures  as only complementary factors for the evaluation of early careers. 

Second, for the entire rank pair $(r_{i}(A),r_{i}(B))$ sample   we calculate a generalized  Kullback-Leibler relative entropy \cite{KL}
\begin{equation}
d(r_{i}(A),r_{i}(B)) \equiv \sum_{i=1}^{S} \Big(r_{i}(A)-r_{i}(B)\Big)\ln \Big(r_{i}(A)/r_{i}(B)\Big)
\label{KL}
\end{equation}
to quantify the relative change in the rankings. 
In each $(r_{i}(A),r_{i}(B))$ scatter plot in Fig. \ref{p4}  we list the normalized value 
\begin{equation}
\overline{d} \equiv \frac{ d(r_{i}(A),r_{i}(B))}{S}
\label{NKL}
\end{equation}
 calculated for all data, and below in parentheses we list  the pair $(\overline{d}_{-},\overline{d}_{+})$, where $\overline{d}_{-}$ is calculated for the lower range $r\in[1,S/2]$ and $\overline{d}_{+}$ is calculated for the upper range $r\in[S/2+1,S]$.

It is important to first note the significant differences in the constituents of the physics dataset with respect to the biology dataset. For the physics dataset, the top $S/2$ careers are all highly prolific scientists, and as a result there is significant rank instability for $r\in[1,S/2]$. However, for the biologist dataset, for which there is a significant difference between dataset [E] and the other biology datasets [F,G,H], the ranking for $r\in[1,S/2]$  is rather stable, while for $r\in[S/2+1,S]$ there is consistently larger instability with  $\overline{d}_{+}>\overline{d}_{-}$. Since the datasets are not well matched, with exception for the 100 top-cited scientists in each, we do not go further into a cross-comparison.

In summary, the largest ``information change'' occurs for the $(Z,g)$ pair, and the least for the $(h,g)$ pair, indicating that the $Z$-index is incorporating additional information into the rankings that $h$ and $g$ are neglecting. This is an important consideration for the large number of careers that are not in the top tier who may experience large
rank instability if information from their entire $c_{i}(r)$ is excluded (say using only $h$) as compared to when it is included (using $h$ in concert with $C$). 

\subsection{Discussion}
%Discussion This should explore the significance of the results of the work, not repeat them. A combined Results and Discussion section is often appropriate. Avoid extensive citations and discussion of published literature.

Here we show that a good indicator should incorporate impact information from the entire $c(r)$ while maintaining simplicity. For $c(r)$ well-described by the DGBD, this would correspond to simply knowing $A_{i}$, $\beta_{i}$, $\gamma_{i}$ (3 parameters) and $N_{i}$ (known for each scientist $i$). Because  $C$ is strongly related to $\beta$ and $h$ through the scaling relation $C\sim h^{1+\beta}$ \cite{RankCitSciRep}, it suffices to know just two of ($\beta$, $C$, and $h$). Since $\beta$ is admittedly tedious to estimate, $C$ and $h$ are the simplest parameters to describe the information contained in $c(r)$.
Hence, the $Z$ measure appeals to the two criteria of comprehensive yet simple (Ockham's razor) by providing a simple geometric representation of the 2-dimensional productivity-impact plane.

Moreover, the $Z$-index is a very simple generalization of the
$h$-index, corresponding for most scientists to a ``renormalization factor'' $R(s)$ that is centered around 1. 
By accounting for the entire citation count $C$, the $Z$-index remedies on of the main weaknesses of $h$, the potentially excessive penalty on a scientist's
high-impact papers, without surrendering the simplicity merit of $h$.
In addition, we have shown that the new $Z$ index is  more
robust towards local changes in the citation profile, an added feature which 
protects against potentially excessive self-citation strategies.
The calculation of the new $Z$ only requires a square root, and could 
be readily and effortlessly incorporated within current major databases, 
such as {\it Google  Scholar}  and {\it ResearcherID.com} profiles, which
already include $h_{i}$ and $C_{i}$.

\section{Conclusions}
%Conclusions
%The main conclusions of the study may be presented in a short Conclusions section, which may stand alone or form a subsection of a Discussion or Results and Discussion section.

The availability of high-resolution career data is opening new avenues in computational social science \cite{CompSocScience}, allowing  insights into the social mechanisms underlying productivity, competition, achievement, and reward \cite{DiffusionRanking,BB2,BB3,StephanIncentives}.
However, in science there has been a proliferation of  indices aimed at measuring simultaneously both productivity and impact, or equivalently, to summarize the entire rank-citation profile $c_{i}(r)$, with a single number. This direction  embraces  simplicity with the potentially paradoxical outcome of  discounting the most notable career achievements.

Here we take a pragmatic approach to measuring the information contained in a scientist's rank-citation profile $c_{i}(r)$ using a 
 2-dimensional representation of  total citations $C_{i}$ and $h_{i}$. 
We propose the 2-component measure $Z$ which is simply a vector norm defined for each  coordinate pair $(h_{i},\sqrt{C_{i}})$ in the $z$-plane.
$Z$ does not discount a scientists's extremely highly cited papers, is less sensitive to local perturbations, and can be readily calculated using $C_{i}$ and $h_{i}$, which are commonly reported in CVs, websites, award applications, and online publication profile services such {\it Google Scholar} and {\it ResearcherID.com}.

It is also important to note that in the practical scenario of career evaluation, if there is going to be a systematic shift towards quantitative measures, then there
should also be measures for the multiple other dimensions of an academic career such as  collaborativity, publication of influential books,  grant writing, teaching awards, mentoring,  administrative leadership, and public and policy-oriented outreach, to name but a few. 
Nevertheless, single-number indicators for productivity and impact are commonly used because of their objective nature, simplicity,
and immediateness. As the scientific labor force continues to grow, and the incentives for producing high-quality scientific products 
continues to change \cite{EconScience}, it will be increasingly important
to understand the evaluation measures underlying career appraisal and their implications 
on the sustainability of career growth  \cite{GrowthCareers,Caution,Predictability2,Reputation}.

\section{Acknowledgments} We thank the anonymous referees for helpful critique. AMP acknowledges COST Action MP0801 and the PNR ``Crisis
  Lab'' project at IMT Lucca.

\bibliography{biblio}

\end{document}